\documentclass[aps,preprint,prd]{revtex4-1}
\usepackage{amssymb}
\usepackage{amsmath}
\usepackage{graphicx}
\usepackage{fancyhdr}
\usepackage{color}
\def\br{\begin{eqnarray}}
\def\er{\end{eqnarray}}
\def\be{\begin{equation}}
\def\ee{\end{equation}}

\def\({\left(}
\def\){\right)}

\newcommand \beq { \begin{eqnarray} }
\newcommand \eeq { \end{eqnarray} }
\newcommand \beqq { \begin{equation} }
\newcommand \eeqq { \end{equation} }

\begin{document}

\title{Flavor Changing Neutral Current Processes in a Reduced Minimal Scalar Sector}

\author{D. Cogollo$^{1}$, Farinaldo S. Queiroz$^{2}$, P. Vasconcelos$^{1}$}

\affiliation{\vspace{0.8cm}
\\
$^{1}$ Departamento de
F\'{\i}sica, Universidade Federal de Campina Grande, Caixa Postal 10071, 58109-970,
Campina Grande, PB, Brazil \vspace{1cm}\\
$^{2}$ Santa Cruz Institute for Particle Physics, University of California, 1156 High St. Santa Cruz, CA 95064, USA.
 \\
} 

\date{\today}

\begin{abstract}

In this work we overhaul previous studies of Flavor Changing Neutral Current processes in the context of the Reduced Minimal 3-3-1 model(RM331). We sift the individual contributions from the CP even scalars and the $Z^{\prime}$ gauge boson using two different parametrizations schemes and compare our results with current measurements. In particular, studying the $B^{0}-\bar{B^{0}}$ meson system we find the most stringent bounds in the literature on this model, namely $M_{Z^{\prime}} \gtrsim 3326$~GeV, $M_{V^{\pm}} \gtrsim 910$~GeV, $M_{U^{++}} \gtrsim 914$~GeV and $m_{h_2^0} \gtrsim 889$~GeV.

\end{abstract}

\maketitle

\section{Introduction}

331 models ~\cite{331-1,331-2} are electroweak gauge extensions of the Standard Model (SM) which might address many questions left out by SM such as the charged quantization \cite{chargequantization}, the number of fermions generations \cite{numerodefamilias}, neutrinos oscillations, the Galactic Center gamma-ray excess \cite{331DM_2} as well as obey the direct detection bounds coming from the XENON100 and LUX experiments \cite{331DM_2,331DM_1,331DM_3,331DM_4,331DM_5,331DM_8}, the dark radiation \cite{331DM_6,331DM_7} observed by Planck \cite{Neff} among others. In the gauge sector, these models add five new gauge bosons that lead to different new physics processes explored in Ref.\cite{gaugebosonsLHC}. In particular, in Ref.\cite{phase331}, it has been shown by the study of phase transition effects that the baryon asymmetry problem might be directly related to the mass of the $Z^{\prime}$ boson. Regarding the scalar sector, the RM331 model \cite{paulodias} is comprised of only two triplet of scalars and hence possesses a reduced scalar sector in comparison with previous ones \cite{331previous}. After the symmetry breaking process, the physical scalar content of the RM331 model is composed of only two neutral CP scalars with lightest one being identified as the SM higgs \cite{explaining1,explaining2}, and a doubly charged one. No singly charged higgs remains in the spectrum differently from other 331 models versions\cite{Martinez:2012ni}.\\

Here we will focus on the FCNC processes that arise at tree level in the RM331 model. In order to cancell the anomalies in this model, two of the quark generations have to live in different representation of $SU(3)_L$ and due to this fact FCNC arise at tree level. In the Standard Model, FCNC processes are forbiden at tree level, they can occur through a weak interaction of second order mediated by the $W$ boson \cite{Albrecht:2012hp}. Many FCNC investigations have been performed in the context of 3-3-1 models \cite{FCNC331,FCNC331_2,FCNC331RM, vicente} (see Ref.\cite{Ferretti:2012se} for supersymmetric models). In particular, a recent analysis of FCNC processes in the RM331 model has been done in Ref.\cite{FCNC331RM}, however, only contributions coming from the $Z^{\prime}$ have been taken into account. Here we plan to expand this investigation calculating the mass difference terms of the $K^{0}-\bar{K^{0}}$, $D^{0}-\bar{D^{0}}$, $B^{0}-\bar{B^{0}}$  meson systems, including both scalars and gauge bosons contributions, as well as those coming from the SM mediated by the $W$ boson  using two particular texture parametrizations for the quark mass matrices, described in \cite{FCNC331,PRDColombia} (parametrization 1) and Ref.\cite{polones} (parametrization 2). Comparing our results with the current experimental limits we derived strong bounds on the mass of the $Z^{\prime}$ boson and the extra CP-even scalar $h_2^0$.  In particular, we find $M_{z^{\prime}} \gtrsim 3326$ Gev and $M_{h_2^0} \gtrsim 889$ GeV. In order to allow the reader to follow our reasoning we will in Sec.\ref{sec1} describe the model. In Sec.\ref{sec2} and Sec.\ref{sec3} we derive the sources of FCNC in the RM331 model coming from the guage and scalar sector respectively. Finally, in Sec.\ref{sec4} we calculate the mass difference of the meson systems and constrain the scale of  symmetry breaking $v_{\chi}$ using two particular texture parametrizations for the quark mass matrices.

\section{THE MODEL}
\label{sec1}

\subsection{Fermions}

The leptonic content of the Reduced Minimal 3-3-1 model (RM331) is comprised of three triplets of type $f^{a}_{L}=(\nu_a, l_a, l^{c}_{a})^{T}_{L} \sim (1,3,0)$, where a=1,2,3 is a family index and the numbers between parentheses represent the fields transformation propierties under the gauge groups $SU(3)_C SU(3)_L U(1)_N$ respectively. In the quark sector, in order to cancel the triangle anomalies one left handed quark family must be arranged in a $SU(3)_L$ triplet whereas the other two come in an anti-triplet representation. In the original version of the RM331 the families of left handed quarks transform as, $Q_{1L}=(u_1, d_1, J_1)^{T}_{L} \sim (3,3,+\frac{2}{3})$, and $Q_{iL}=(d_i,-u_i, J_i)^{T}_{L} \sim (3,3^*,-\frac{1}{3})$ \cite{paulodias}, where $i=2,3$ is a family index and $J_1,J_2$ and $J_3$ are new heavy quarks characteristic of these theories. However, it has been shown in Ref.\cite{FCNC331RM} that this representation choice must be ruled out because it induces FCNC contributions that exceed the current experimental limits. The only way to evade these bounds is setting the $Z^{\prime}$ mass above the ~$100$~TeV scale. Albeit, the RM331 model is valid only up to $\approx$ 5 TeV, thus such a high mass for the $Z^{\prime}$ boson is prohibited, leading us to the conclusion that the original representation of the left handed quarks families is excluded.  For this reason we will adopt that the left handed quark families transform as in references \cite{vicente, landau}, $Q_{3L}=(u_3, d_3, J_3)^{T}_{L} \sim (3,3,+\frac{2}{3}), Q_{iL}=(d_i, -u_i, J_i)^{T}_{L} \sim (3,3^*,-\frac{1}{3})$, where $i=1,2$. With this quark representation the theory is anomaly free as well, since the requirement that the number of triples be equal to the number of anti-triplets is respected. Notice that because of colors, one quark generation actually counts as three. Hence three lepton generation plus one quark generation results in six triplets, whereas the two quark generations that live in the anti-triplet representation accounts as 6 anti-triplets. Thus the triangle anomalies are cancelled too. As for the Right handed quarks, suffice it to say that are singlets of the $SU(3)_{L}$ group.\\
\subsection*{Scalars and Gauge Bosons}
The scalar content of the model is comprised of two triplets, $\rho=(\rho^{+},\rho^{0}, \rho^{++})^{T} \sim (1,3,1)$, and $\chi=(\chi^{-}, \chi^{--}, \chi^{0})^{T} \sim (1,3,-1)$. The neutral scalars $\rho^0$ and $\chi^0$ can develop a vev different from zero, as we promptly do by shifting these fields as
\begin{equation}
\rho^0, \chi^0 \rightarrow \dfrac{1}{\sqrt{2}}(v_{\rho,\chi} + R_{\rho,\chi} + iI_{\rho,\chi}),
\end{equation}

which suffice to break correctly the $SU(3)_C \times SU(3)_L \times U(1)_N$ symmetry into the $SU(3)_C \times U(1)_{QED}$ and generates the correct masses of all fermions, including neutrinos, and gauge bosons. The symmetry breaking follows the pattern \be {\rm SU(3)}_L\otimes{\rm
U(1)}_N \stackrel{\langle\chi^0\rangle} \longrightarrow {\rm
SU(2)}_L\otimes{\rm U(1)}_Y \stackrel{\langle \rho^0\rangle}
\longrightarrow {\rm U(1)}_{\rm EM},\ee
and so $v_\chi \gg v_\rho$. The most general, renormalizable, gauge and Lorentz invariant scalar potential of the RM331 model is
\begin{eqnarray}
V(\chi, \rho) &=& \mu_1^2 \rho^{\dagger}\rho + \mu_2^2 \chi^{\dagger}\chi + \lambda_1(\rho^{\dagger}\rho)^2 + \lambda_2 (\chi^{\dagger}\chi)^2 \nonumber\\
&&+ \lambda_3 (\rho^{\dagger}\rho)(\chi^{\dagger}\chi) + \lambda_4(\rho^{\dagger}\chi)(\chi^{\dagger}\rho).
\end{eqnarray}

After the symmetry breaking the mass matrix of the neutral CP scalars in the basis $(R_\chi\,,\,R_\rho)$ was found to be
\begin{equation}
m^2_0=\frac{v^2_\chi}{2}
\begin{pmatrix}
2\lambda_2\ & \lambda_3 t  \\
\lambda_ 3 t& 2\lambda_1 t^2 
\end{pmatrix},
\label{neutralmassmatrix}
\end{equation} with $t=\frac{v_\rho}{v_\chi}$. The diagonalization of this matrix leads to the following eigenstates in the limit ($v_\chi \gg v_\rho$), 
\begin{equation}
h_1^0 = c_\beta R_{\rho} - s_\beta R_\chi, \quad h_2^0 = c_\beta R_{\chi} + s_\beta R_\rho,
\label{neutralstates}
\end{equation}and masses
\begin{equation}
m^2_{h_1^0} = \left(\lambda_1 - \dfrac{\lambda_3^2}{4\lambda_2}\right)v_\rho^2, \quad
m^2_{h_2^0} = \lambda_2 v_\chi^2 + \dfrac{\lambda_3^2}{4\lambda_2} v_\rho^2,
\label{massadosneutros}
\end{equation}
with $\lambda_1, \lambda_2 > 0 $, $c_\beta \equiv cos_\beta \approx 1 - \dfrac{\lambda_3^2}{8\lambda_2^2}\dfrac{v_\rho^2}{v_\chi^2}$ and $s_\beta \equiv sin_\beta \approx \dfrac{\lambda_3}{2\lambda_2}\dfrac{v_\rho}{v_\chi}$. The condition $v_\chi \gg v_\rho$ implies that $m^2_{h_1^0} \ll m^2_{h_2^0}$ and $c_{\beta} \gg s_{\beta}$. $h_1^0$ is identified as SM higgs boson, its interactions with standard particles are shown in the Table \ref{tableh1} (where q=u,c,b. $q^{\prime}$=d,s,t). In the limit when $\sin_{\beta} \rightarrow 0$ and $\cos_{\beta} \rightarrow 1$ we find $h_1^0 \equiv h$, i.e, we recover the standard Higgs, as it must be. With the ATLAS/CMS Collaborations \cite{atlas,cms} announcement of the BEH SM scalar with a mass of $125$~GeV, we enforce the relation $\lambda_1 - \dfrac{\lambda_3^2}{4\lambda_2} \simeq \dfrac{1}{4}$, in Eq (\ref{massadosneutros}). This constraint will be obeyed throughout this work because it guarantees that the mass of $h_1^0$ be $125$~GeV. Furthermore, since $M^2_{W^{\pm}} = \dfrac{g^2 v^{2}_{\rho}}{4}$, we have used $v_{\rho} \approx 246$ GeV.\\ 
Regarding the pseudo-scalars, $I_\rho$  and $I_\chi$, they are both Goldstone bosons eaten by the gauge bosons $Z_\mu$ and $Z^{\prime}_\mu$, respectively, whereas the charged scalars are absorbed by the gauge bosons $W^{+}$ and $V^{+}$, and as for the doubly charged ones, one of them is a Goldstone eaten by $U^{++}$ and the other remains in the physical spectrum with mass $M^2_{h^{++}}=\dfrac{\lambda_4}{2}(v^2_{\chi}+v^2_{\rho})$. Summarizing, the masses of the five new gauge bosons are given by,

$$ m^2_{Z^{\prime}}=\frac{g^2 c^2_W}{3(1-4 s^2_W)}v^2_\chi\ , \quad M^2_{V{^\pm}} = \dfrac{g^2 v^2_{\chi}}{4} \quad,  $$

$$ M^2_{U{^{\pm \pm}}} = \dfrac{g^2 (v^2_{\rho} + v^2_{\chi})}{4}. $$

Further we will show the neutral currents of the model because from these we will derive the FCNC processes.

\begin{table}
\centering
\caption{Higgs-like ($h_1^0$) Standard interactions.}
\label{tablecv}
\begin{tabular*}{\columnwidth}{@{\extracolsep{\fill}}llllll@{}}
\hline
\multicolumn{1}{@{}l}{Interactions} & Couplings\\
\hline
$\overline{l}lh_1^0$ & $\dfrac{m_l}{v_{\rho}}\left(c_\beta - \frac{v_\rho}{v_\chi}s_\beta\right)$ \\
$\overline{q}qh_1^0$   & $\dfrac{m_q}{v_{\rho}}c_{\beta}$ \\ 
$\overline{q'}q'h_1^0$ & $\dfrac{m_q'}{v_{\rho}}\left(c_\beta - \frac{v_\rho}{v_\chi}s_\beta\right)$ \\
$W^+W^-h_1^0$          &  $\frac{1}{2} g^2v_{\rho}c_{\beta}$\\
$ZZh_1^0$ 				   & $\frac{1}{4} g^2v_{\rho}\sec^{2}_{\theta_W}c_{\beta}$\\\hline
\end{tabular*}
\label{tableh1}
\end{table}

\section{Neutral currents via gauge boson exchanges}
\label{sec2}

The neutral currents of the standard down quarks mediated by the extra neutral gauge boson $Z^{\prime}$ comes from the following lagrangian \cite{paulodias},

\beq
{\cal L}^{NC}_{Z^{\prime} ,d}& =& \left( \dfrac{g}{2C_{W}}(\bar D^{\prime} \gamma_\mu(1-\gamma_5)Y_{Z^{\prime}}D^{\prime}\right) 
Z^{\prime}_{\mu},
\label{fcncz} 
\eeq
where $Y_{Z^{\prime}}=\frac{1}{\sqrt{12h_W}}diag(1-2s^2_W,1-2s^2_W , 1)$, $h_{W}=1-4S^{2}_{W}$, and $D^{\prime}=(d^{\prime}_{1}, d^{\prime}_{2}, d^{\prime}_{3})^T$ is the flavor basis of down quarks. The structure of the function $Y_{Z^{\prime}}$ highlights that the universality of the interactions mediated by the $Z^{\prime}$ boson has been lost. Therefore, we may rewrite Eq.(\ref{fcncz}) explicitly as,

\begin{eqnarray}
{\cal L}^{NC}_{Z^{\prime} ,d}&=&\dfrac{g}{C_{W}\sqrt{12h_{W}}}(\bar{d^{\prime}}_{3L}\gamma^{\mu}d^{\prime}_{3L}+\bar{d^{\prime}}_{iL}\gamma^{\mu}(1-2S^{2}_{W})d^{\prime}_{iL})Z^{\prime}_{\mu},
\label{fcnc1}
\end{eqnarray}and after rearranging the terms as, 
\begin{eqnarray}
{\cal L}^{NC}_{Z^{\prime} ,d}&=&\dfrac{g}{C_{W}\sqrt{12h_{W}}}
\left[\sum_{a=1}^{3}(\bar{d^{\prime}}_{aL}\gamma^{\mu}(1-2S^{2}_{W})d^{\prime}_{aL}+\bar{d^{\prime}}_{3L}\gamma^{\mu}(2S^{2}_{W})d^{\prime}_{3L}\right]Z^{\prime}_{\mu},
\label{fcnc2}
\end{eqnarray}
with $a=1,2,3$ being the family index. From Eq.\eqref{fcnc2} we can observe that only the second term induces FCNCs at tree level. Similarly, the neutral currents of the up quarks mediated by the extra neutral gauge boson $Z^{\prime}$ are :
\begin{eqnarray}
{\cal L}^{NC}_{Z^{\prime} ,u}&=&\dfrac{g}{C_{W}\sqrt{12h_{W}}}
(\sum_{a=1}^{3}(\bar{u^{\prime}}_{aL}\gamma^{\mu}(1-2S^{2}_{W})u^{\prime}_{aL}+\bar{u^{\prime}}_{3L}\gamma^{\mu}(2S^{2}_{W})u^{\prime}_{3L})Z^{\prime}_{\mu},
\label{fcnc3}
\end{eqnarray}with the last term being the only source of FCNC at tree level. It is well known that the flavor and physical quark bases are related through,
\begin{equation}
\begin{pmatrix}
u\\
c\\
t
\end{pmatrix}_{L,R}
=
V_{L,R}^{u}
\begin{pmatrix}
u^{\prime}\\
c^{\prime}\\
t^{\prime}
\end{pmatrix}_{L,R},
\begin{pmatrix}
d\\
s\\
b
\end{pmatrix}_{L,R}
=
V_{L,R}^{d}
\begin{pmatrix}
d^{\prime}\\
s^{\prime}\\
b^{\prime}
\end{pmatrix},
\label{misturaquarksMP}
\end{equation}

where $V_{L,R}^{u}$ and $V_{L,R}^{d}$ are $3\times 3$ unitary matrices which diagonalize the mass matrices for up and down quarks. Applying the transformations determined in Eq.\eqref{misturaquarksMP} into Eqs.\eqref{fcnc2}-\eqref{fcnc3} we obtain the interactions among the physical quarks and the $Z^{\prime}$ boson (appendix~{\bf A}) that contribute to the mass difference terms of the meson systems as we shall see further.

\section{Neutral Currents via Scalar Exchange}
\label{sec3}

Now we will devote this section to obtain the sources of FCNC coming from the scalar sector, but in order to do so we first need to get the mass matrices of the quarks, which are derived from a combination of the renormalizable Yukawa lagrangian plus effective dimension five operators given by,

\begin{eqnarray}
&&
 \lambda^d_{3a}\bar Q_{3L}\rho d_{aR} + \frac{\lambda^d_{ia}}{\Lambda}\varepsilon_{nmp}\left(\bar Q_{iLn}\rho_m\chi_p\right)d_{aR} + \nonumber \\
 && \lambda^u_{ia}\bar Q_{iL}\rho^* u_{aR} + \frac{\lambda^u_{3a}}{\Lambda}\varepsilon_{nmp}\left(\bar Q_{3Ln}\rho^*_m\chi^*_p\right)u_{aR} + H.C.
 \label{quarksmassterms}
\end{eqnarray}

with $i=1,2$ and $\Lambda=4-5 $TeV~\cite{landau} being the highest energy scale where the model is found to be valid. From Eq.(\ref{quarksmassterms}) we found the quark mass matrices in the flavor basis $U^{\prime}$ and $D^{\prime}$ to be respectively:

\begin{equation}
m^u \approx
\begin{pmatrix}
\frac{-\lambda^u_{11}v_\rho}{\sqrt{2}}  & \frac{-\lambda^u_{12}v_\rho}{\sqrt{2}} & \frac{-\lambda^u_{13}v_\rho}{\sqrt{2}}   \\
\frac{-\lambda^u_{21}v_\rho}{\sqrt{2}}  & \frac{-\lambda^u_{22}v_\rho}{\sqrt{2}} & \frac{-\lambda^u_{23}v_\rho}{\sqrt{2}}   \\
\frac{\lambda^u_{31}v_\rho}{2} & \frac{\lambda^u_{32}v_\rho}{2} & \frac{\lambda^u_{33}v_\rho}{2}
\end{pmatrix},
m^d \approx 
\begin{pmatrix}
\frac{\lambda^d_{11}v_\rho}{2}  & \frac{\lambda^d_{12}v_\rho}{2} & \frac{\lambda^d_{13}v_\rho}{2}   \\
\frac{\lambda^d_{21}v_\rho}{2}  & \frac{\lambda^d_{22}v_\rho}{2} & \frac{\lambda^d_{23}v_\rho}{2}   \\
\frac{\lambda^d_{31}v_\rho}{\sqrt{2}} & \frac{\lambda^d_{32}v_\rho}{\sqrt{2}} & \frac{\lambda^d_{33}v_\rho}{\sqrt{2}}
\end{pmatrix}
\label{massmatrixupquarks},
\end{equation}where the approximation $v_{\chi} \approx \Lambda$ has been used. Now that we have obtained the mass matrices in Eq.\eqref{massmatrixupquarks}, we are able to write the interactions among the flavor quarks eigenstates and the physical scalars $h_{1,2}^0$, wich are also derived from \eqref{quarksmassterms},
\begin{equation}\label{lagrangian}
\mathcal{L} = \overline{U^{\prime}}_L\Gamma_1^uU^{\prime}_Rh_1^0 + \overline{U^{\prime}}_L\Gamma_2^uU^{\prime}_Rh_2^0
+ \overline{D^{\prime}}_L\Gamma_1^dD^{\prime}_Rh_1^0 + \overline{D^{\prime}}_L\Gamma_2^dD^{\prime}_Rh_2^0 + h.c,
\end{equation} with,
\begin{eqnarray}
\Gamma_1^u &=& \dfrac{m^{u}}{v_{\rho}}\cos_\beta - \dfrac{\sin_\beta}{v_{\chi}}
\left(\begin{array}{ccc} 
0 & 0 & 0\\
0 & 0 & 0 \\
m^{u}_{31} & m^{u}_{32} & m^{u}_{33} \\ 
\end{array}\right), 
\label{gu1}
\end{eqnarray}
\begin{eqnarray}
\Gamma_1^d &=& \dfrac{m^{d}}{v_{\rho}}\cos_\beta - \dfrac{\sin_\beta}{v_{\chi}} 
\left(\begin{array}{ccc} 
m^{d}_{11} & m^{d}_{12} & m^{d}_{13}\\
m^{d}_{21} & m^{d}_{22} & m^{d}_{23} \\
0 & 0 & 0 \\ 
\end{array}\right),
\label{gd1}
\end{eqnarray}
\begin{eqnarray}
\Gamma_2^u &=& \dfrac{m^{u}}{v_{\rho}}\sin_\beta + \dfrac{\cos_\beta}{v_{\chi}}
\left(\begin{array}{ccc} 
0 & 0 & 0\\
0 & 0 & 0 \\
m^{u}_{31} & m^{u}_{32} & m^{u}_{33} \\ 
\end{array}\right),
\label{gu2}
\end{eqnarray}
\begin{eqnarray}
\Gamma_2^d &=& \dfrac{m^{d}}{v_{\rho}}\sin_\beta + \dfrac{\cos_\beta}{v_{\chi}}
\left(\begin{array}{ccc} 
m^{d}_{11} & m^{d}_{12} & m^{d}_{13}\\
m^{d}_{21} & m^{d}_{22} & m^{d}_{23} \\
0 & 0 & 0 \\ 
\end{array}\right).
\label{gd2}
\end{eqnarray}
It is important to make a few remarks concerning the Eq.\ref{lagrangian}.
\begin{itemize}

\item Firstly, after applying the transformations given in Eq.\eqref{misturaquarksMP} into Eq.\eqref{lagrangian} only the first terms of the $\Gamma^{u,d}$ matrices, wich are proportional to the quark mass matrices $m^{u,d}$, are diagonalized. Meanwhile, the second terms will induce non-diagonal interactions mediated by these scalars.

\item Secondly, the origin of these second terms in Eqs.(\ref{gu1})-(\ref{gd2}) are related to the fact that the {\it three left handed quark generations do not transform in the same way and it is precisely for this particularity those scalars may mediate FCNC processes} \cite{FCNC331,vicente}. These second terms are all suppressed by the scale of symmetry breaking of the model, $v_{\chi}$. In particular, the second terms in Eqs.(\ref{gu1})-(\ref{gd1}) which refer to SM higgs mediated processes, are extremely suppressed once we are taking the limit $\sin_{\beta} \rightarrow 0$ and $v_{\chi} \gg v_{\rho}$.

Once we have discussed the physical effects of Eq.(\ref{lagrangian}), we will write down the terms that induce FCNC after applying the transformations given in Eq.\eqref{misturaquarksMP} as discussed above.

\end{itemize}

\begin{equation}
\mathcal{L^{FCNC}}_{h_1^0} = -\sin_\beta\bar{U}_{L}K^{U}U_{R}h_1^0-\sin_\beta\bar{D}_{L}K^{D}D_{R}h_1^0 + H.C,
\label{gu5}
\end{equation}
\begin{equation}
\mathcal{L^{FCNC}}_{h_2^0} = \cos_\beta\bar{U}_{L}K^{U}U_{R}h_2^0
+\cos_\beta\bar{D}_{L}K^{D}D_{R}h_2^0 + H.C, 
\label{gu6}
\end{equation}where,
\begin{eqnarray}
K^{U}=V^{u}_{L} 
\left(\begin{array}{ccc} 
0 & 0 & 0\\
0 & 0 & 0 \\
\dfrac{m^{u}_{31}}{v_{\chi}} & \dfrac{m^{u}_{32}}{v_{\chi}} & \dfrac{m^{u}_{32}}{v_{\chi}} \\ 
\end{array}\right)(V^{u}_{R})^{\dagger},
\label{gu7}
\end{eqnarray}
\begin{eqnarray}
K^{D}=V^{d}_{L} 
\left(\begin{array}{ccc} 
\dfrac{ m^{d}_{11} }{ v_{\chi} }  & \dfrac{ m^{d}_{12}}{v_{\chi}} & \dfrac{m^{d}_{13}}{v_{\chi}}\\
\dfrac{m^{d}_{21}}{v_{\chi}} & \dfrac{m^{d}_{22}}{v_{\chi}} & \dfrac{m^{d}_{23}}{v_{\chi}}\\
0 & 0 & 0 \\ 
\end{array}\right)(V^{d}_{R})^{\dagger}. 
\label{gu8}
\end{eqnarray}

We have shown in (appendix~{\bf A}) the explicit form of Eqs.(\ref{gu5})-(\ref{gu6}).

\section{Mass difference of the neutral Mesons Systems}
\label{sec4}
Now that we have derived the FCNC lagrangians mediated by the $Z^{\prime}$ boson and the CP-even scalars $h_1^0$ and $h_2^0$, we can move forward  and calculate the RM331 contributions to the mass difference of the mesons systems $K^{0}-\bar{K^{0}}$, $D^{0}-\bar{D^{0}}$, $B^{0}-\bar{B^{0}}$. Hereafter  {\it we will include the SM contributions and set stringent bounds on the masses of the $Z^{\prime}$ and $h_2^0$ bosons}. We will start studying the $B^{0}-\bar{B^{0}}$ system where the stringent bounds were found

\subsection{$B^{0}-\bar{B^{0}}$ System}

We first consider the $Z^{\prime}$ contribution. The effective $Z^{\prime}$ lagrangian that induces $B^{0}_{d} \rightarrow \bar{B^{0}}_{d}$ transitions is obtained straightforwardly from Eq.\eqref{BB},

\begin{eqnarray}
\ensuremath{\mathcal{L}}^{B^{0}-\bar{B^{0}}}_{Z'\ eff} & = & \frac{4 \sqrt{2} G_F S^4_W}{(3h_W)}\frac{M_{Z}^{2}}{M_{Z^{\prime}}^{2}}[(V_{L}^{d})_{13}^{\ast}(V_{L}^{d})_{33}]^{2}[\bar{d}_{3L}\gamma_{\mu}d_{1L}]^{2}.\nonumber \\
\label{effectiveLB}
\end{eqnarray}

Here we assume that the mixing angle among the physical eigenstates $Z_1$ and $Z_2$ is negligible and therefore $Z_{1}=Z$ and $Z_{2}=Z^{\prime}$. That being said, the RM331 contribution to $(\Delta m_{B})_{Z^{\prime}}$ is given by

\begin{eqnarray}
(\Delta m_{B})_{Z^{\prime}} & = & <\bar{B^ {0}}|\ensuremath{\mathcal{L}}^{B_{0}-\bar{D_{0}}}_{Z'\ eff}|B^ {0}>\nonumber \\
&=&\frac{4 \sqrt{2} G_F S^4_W}{(3h_W)}\frac{M_{Z}^{2}}{M_{Z^{\prime}}^{2}}[(V_{L}^{d})_{13}^{\ast}(V_{L}^{d})_{33}]^{2}<\bar{B^ {0}}|(\bar{d_3}d_1)_{V-A}^2|K^ {0}>,
\label{massdifBZlinha}
\end{eqnarray}where

\begin{equation}
<\bar{B^ {0}}|(\bar{d_3}d_1)_{V-A}^2|B^ {0}>=\frac{M_BB_{B}f_{B}^{2}}{3}
\label{approx}
\end{equation} 

according to the vacuum insertion approximation \cite{vacum}. $B_{B}$ and $f_{B}$ are the bag parameter and the decay constant of the meson respectively. For Next leading order QCD corrections of Eq.\eqref{approx} the reader can see ~\cite{etaparameter}. 

As aforementioned the Higgs as well as the heavy scalar $h^{0}_{2}$ mediate FCNC processes. The effective Lagrangian that induces $B^{0}_{d} \rightarrow \bar{B^{0}}_{d}$ transitions is obtained straightforwardly from Eq.\eqref{BBscalar} in the Appendix as follows,

\begin{eqnarray}
\ensuremath{\mathcal{L}}^{B_{0}-\bar{B_{0}}}_{h1,h2\ eff} & = & \dfrac{\sin_{\beta}^{2}}{m^{2}_{h_1^0}}\left[\left[K^{D}_{3,1}+(K^{D}_{1,3})^{\ast}\right]^{2}(\bar{d_3}d_1)^{2}+\left[K^{D}_{3,1}-(K^{D}_{1,3})^{\ast}\right]^{2}(\bar{d_3}\gamma_5d_{1})^{2}\right]\nonumber \\
& &+\dfrac{\cos_{\beta}^{2}}{m^{2}_{h_2^0}}\left[\left[K^{D}_{3,1}+(K^{D}_{1,3})^{\ast}\right]^{2}(\bar{d_3}d_1)^{2}+\left[K^{D}_{3,1}-(K^{D}_{1,3})^{\ast}\right]^{2}(\bar{d_3}\gamma_5d_{1})^{2}\right].
\label{effectiveSLB}
\end{eqnarray}
Defining,
\begin{equation}
(K^{U,D}_{i,j})^{\pm}=K^{U,D}_{i,j} \pm (K^{U,D}_{j,i})^{\ast}.
\end{equation}we find,
\begin{eqnarray}
(\Delta m_{B})_{h_1^0,h_2^0} & = & <\bar{B^ {0}}|\ensuremath{\mathcal{L}}^{B_{0}-\bar{B_{0}}}_{h_1^0,h_2^0\ eff}|B^ {0}>\nonumber\\
& = & \lbrace\dfrac{\sin_{\beta}^{2}}{4m^{2}_{h_1^0}}\left[-(K^{D+}_{3,1})^{2}(1-\frac{M_{B}^{2}}{(m_{d}+m_{b})^{2}})+(K^{D-}_{3,1})^{2}(1-11\frac{M_{B}^{2}}{(m_{d}+m_{b})^{2}})\right]\nonumber\\
&+& \dfrac{\cos_{\beta}^{2}}{4m^{2}_{h_2^0}}\left[-(K^{D+}_{3,1})^{2}(1-\frac{M_{B}^{2}}{(m_{d}+m_{b})^{2}})+(K^{D-}_{3,1})^{2}(1-11\frac{M_{B}^{2}}{(m_{d}+m_{b})^{2}})\right]\rbrace\nonumber\\
& & \times<\bar{B^ {0}}|(\bar{d_3}d_1)_{V-A}^2|B^ {0}>.
\label{massdifBh-1h1h2}
\end{eqnarray}with, 
\begin{eqnarray}
<\bar{B^ {0}}|(\bar{d_3}d_1)^{2}|B^{0}> & = & -\frac{1}{4}\left[1-\frac{M_{B}^{2}}{(m_{b}+m_{d})^{2}}\right]<\bar{B^ {0}}|(\bar{d_3}d_1)_{V-A}^2|B^ {0}>,\nonumber\\
<\bar{B^ {0}}|(\bar{d_3}\gamma_5d_1)^{2}|B^{0}> & = & \frac{1}{4}\left[1-11\frac{M_{B}^{2}}{(m_{b}+m_{d})^{2}}\right]<\bar{B^ {0}}|(\bar{d_3}d_1)_{V-A}^2|B^ {0}>,
\label{scalarboson}
\end{eqnarray}
in agreement with \cite{vicente}. 

Combining Eq.(\ref{massdifBZlinha}) and (\ref{massdifBh-1h1h2}) we have the contributions coming from the RM331 model. The latter should complemented with the SM one given by,
\begin{eqnarray}
(\Delta m_{B})_{SM} & = & \dfrac{G_{f}^{2}M_{W}^{2}}{12\pi^2}S_{0}(x_{t})[(V_{CKM})_{td}^{\ast}(V_{CKM})_{tb}]^2<\bar{B^ {0}}|(\bar{d_2}d_1)_{V-A}^2|B^ {0}>,
\end{eqnarray}
where $x_{t}=\frac{m_{t}^{2}}{M_{W}^{2}}$ and $S_{0}(x_{t}) \approx 0.784x_{t}^{0.76}$ \cite{vicente}. 

At this moment it is important to emphasize:

\begin{itemize}

\item We are forcing $M_{h_1^0} = 125$~GeV because $h_1^0$ is recognized as the SM Higgs in our model. In this case $\sin_{\beta} \rightarrow 0$. Therefore the SM Higgs does not  mediate relevant FCNC processes in accordance with the current data \cite{atlas}.
\item The total contribution of RM331 model $(\Delta m_{B})_{RM331}=(\Delta m_{B})_{Z^{\prime}}+(\Delta m_{B})_{h_1^0}+(\Delta m_{B})_{h_2^0}+(\Delta m_{B})_{SM}$ depends primarily on the scale of symmetry breaking of the model ($v_{\chi}$).

\item Albeit, the total contribution $(\Delta m_{B})_{RM331}$ also depends on the matrix elements that relate the flavor and physical quarks basis for one side as well as on the mass matrix in the flavour basis of the quarks as we can see in Eq.\eqref{massdifBZlinha} and Eq.\eqref{massdifBh-1h1h2}.
\end{itemize}

That being said, we obtain our results using two different parametrizations of Fritzsch type for the quark mass matrices \cite{PRDColombia}-\cite{polones} and their respective unitary matrices $V^{u,d}_{L,R}$ that diagonalized them. Hereafter we will name the parametrization from Ref.\cite{PRDColombia} as parametrization 1 (see appendix~{\bf B}), and parametrization 2 the one used in Ref.\cite{polones} (see appendix~{\bf C}).

\begin{figure}[!t]
\mbox{\includegraphics[scale=0.9]{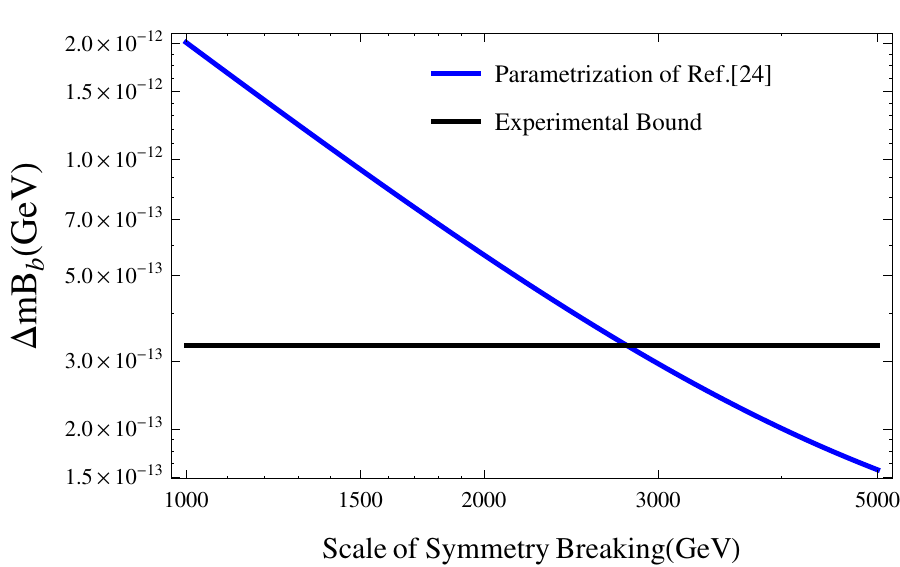}
\includegraphics[scale=0.9]{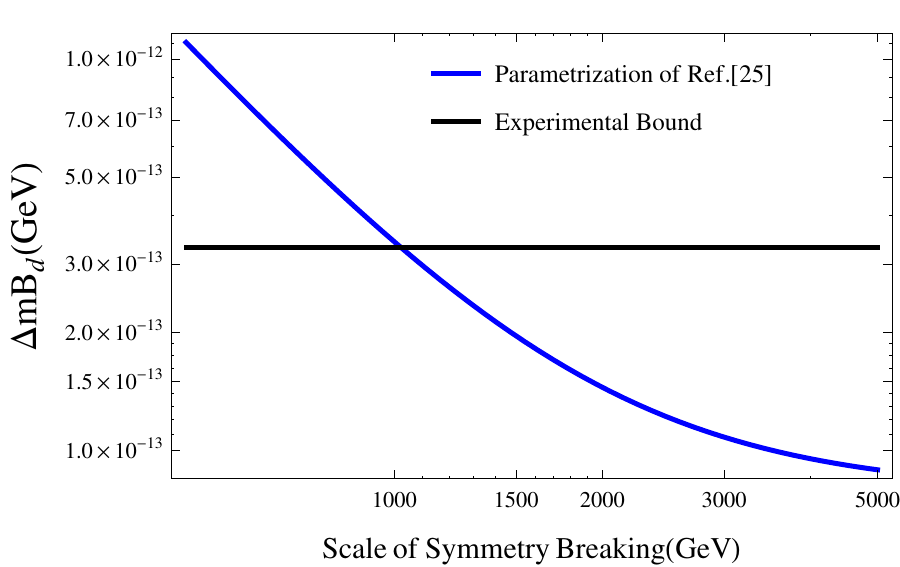}}
\caption{Total contribution from the RM331 model to $(\Delta m_{B})=(\Delta m_{B})_{RM331}=(\Delta m_{B})_{Z^{\prime}}+(\Delta m_{B})_{h_1^0}+(\Delta m_{B})_{h_2^0}+(\Delta m_{B})_{SM}$ as function of the scale of symmetry breaking using 
parametrization described in Ref.\cite{PRDColombia} ({\it left panel}) and the texture parametrization of Ref.\cite{polones} ({\it right panel}). The horizontal black line refers to the current experimental limit. Thus we conclude that $v_{\chi} \gtrsim 2786$~GeV (with Ref.\cite{PRDColombia} textures), which implies that $M_{Z^{\prime}} \gtrsim 3326$~GeV, $M_{V^{\pm}} \gtrsim 910$~GeV, $M_{U^{++}} \gtrsim 914$~GeV, $m_{h2} \gtrsim 889$~GeV. Using the parametrization of Ref.\cite{polones} we find $v_{\chi} \gtrsim 1023$~GeV which translates into $M_{Z^{\prime}} \gtrsim 1221 $~GeV, $M_{V^{\pm}} \gtrsim 334$~GeV, $M_{U^{++}} \gtrsim 343$~GeV, $m_{h2} \gtrsim 345$~GeV. We have used $S_W^2=0.231$.}
\label{fig3}
\end{figure}

In FIG.\ref{fig3} we show the total contribution from the RM331 model $(\Delta m_{B})_{RM331}=(\Delta m_{B})_{Z^{\prime}}+(\Delta m_{B})_{h_1^0}+(\Delta m_{B})_{h_2^0}+(\Delta m_{B})_{SM}$ in terms of $v_{\chi}$. We plotted the results for the two parametrizations. In left panel we used parametrization 1 described in Ref.\cite{PRDColombia}, whereas in right panel we applied parametrization 2 studied in Ref.\cite{polones}. We remark on the two parametrizations that induce two distinct results and consequently different bounds on $v_{\chi}$. Comparing our results with the current experimental limit on the mass difference of the B meson system namely, $(\Delta m_{B_d}) \leqslant 3.33 \times 10^{-13}$~GeV, and using the values $m_{B_d} = 5279.5 \ \mbox{MeV},
\sqrt{B_{B_d}}f_{B_d} = 208\ \mbox{MeV}$, according to \cite{PDG}, we obtain $v_{\chi} \gtrsim 2786$~GeV (parametrization 1) and $v_{\chi} \gtrsim 1023$~GeV (parametrization 2).

{\it These bounds on the scale of symmetry breaking of the model are relevant because they have a direct impact on the masses of the gauge bosons and the heavy Higgs}. In the table \ref{table2} we have summarized our findings. Similar bounds were found in a recent study concerning the $(g-2)_{\mu}$ anomaly in Ref.\cite{g-2331}.

\begin{table}[t]
{\bf Bounds from FCNC in the RM331 Model}\\
\begin{tabular}{|c|c|}
\hline
Parametrization 1 (Ref.\cite{PRDColombia}) & $M_{Z^{\prime}} \gtrsim 3.326$~TeV , $M_{V^{\pm}} \gtrsim 0.910$~TeV\\ & $M_{U^{++}} \gtrsim 0.914$~TeV , $m_{h_2^0} \gtrsim 0.889$~TeV\\
\hline
Parametrization 2 (Ref.\cite{polones}) & $M_{Z^{\prime}} \gtrsim 1.221$~TeV, $M_{V^{\pm}} \gtrsim 0.334$~TeV\\ & $M_{U^{++}} \gtrsim 0.343$~TeV, $m_{h_2^0} \gtrsim 0.345$~TeV\\
\hline
\end{tabular}
\caption{Summary of bounds derived from $B^{0}-\bar{B^{0}}$ system with $S_W^2= 0.231$. Investigating FCNC processes in the RM331 model, we have set the most stringent bounds on the mass spectrum in the literature.}
\label{table2}
\end{table} 

\subsection{$K^{0}-\bar{K^{0}}$ and $D^{0}-\bar{D^{0}}$ Systems}

As for the other meson systems suffice it to say that we did not find any relevant bound on the scale of symmetry breaking. For instance, comparing our results of $(\Delta m_{D})_{RM331}$ with the current experimental limit, which reads $(\Delta m_{D}) \leqslant 9.478 \times 10^{-15}$~GeV, we set $v_{\chi} \gtrsim 18$~GeV in the parametrization of Ref.\cite{PRDColombia}, which is completely irrelevant. The explicit form of $(\Delta m_{D})_{RM331}$ and $(\Delta m_{K})_{RM331}$ are shown in appendix~{\bf D}

\section{Conclusions}

We have sifted the sources of FCNC in the RM331 model. We have concluded that the third left handed quark generation should transform in the triplet representation of $SU(3)_L$, whereas the first two in the anti-triplet one, otherwise huge FCNC contributions arise exceeding the current experimental limits, in agreement with previous results, and the only way to circumvent this bound is by setting the $Z^{\prime}$ mass above ~$100$~TeV  which is by far beyond the perturbative limit of the model $\sim 5$TeV. Furthermore, we noticed that besides the $Z^{\prime}$ boson, the CP-even scalars of the model may mediate sizeable FCNC processes as well. Moreover, we have computed the mass difference terms for $D^{0}-\bar{D^{0}}$, $K^{0}-\bar{K^{0}}$, and $B^{0}-\bar{B^{0}}$ using two different parametrizations schemes for mass matrices of the quarks and unitary matrices $V^{u,d}_{L,R}$, with the $B^{0}-\bar{B^{0}}$ system offering the only relevant bound on the scale of symmetry breaking of the model. Comparing our results with the current experimental limits on $\Delta m_{B}$ we have set $v_{\chi} \gtrsim 2786$~GeV using the texture parametrization of Ref.\cite{PRDColombia,FCNC331} and $v_{\chi} \gtrsim 1023$~GeV for the parametrization of Ref.\cite{polones}. Our results are summarized in table \ref{table2}. In particular, using parametrization 1 we have found this model to be consistent with FCNC limits if,

\begin{itemize}
\item $M_{Z^{\prime}} \gtrsim 3326$~GeV,
\item $M_{V^{\pm}} \gtrsim 910$~GeV,
\item $M_{U^{++}} \gtrsim 914$~GeV,
\item $m_{h_2^0} \gtrsim 889$~GeV.
\end{itemize} 

It is important to point out that different parametrization schemes in the quark sector might affect our conclusions.

\begin{acknowledgements}
The authors thank Vicente Pleitez for valuable discussions and comments as well as Paulo Rodrigues and Paulo Rogerio for clarifying some issues in Ref.\cite{paulodias}. The authors also thank Patrick Draper, William Shepherd and Alex Dias. DC is partly supported by Conselho Nacional de Desenvolvimento Cientifico e Tecnologico (CNPq) Grant 484157/2013-2,
FSQ by Department of Energy Award SC0010107 and CNPq, and PV by  Coordena\c{c}\~ao de Aperfei\c{c}oamento de Pessoal de N\'{\i}vel Superior.
\end{acknowledgements}
  
\appendix
\section{}
\label{appendix1}

Interactions among the physical standard quarks and the $Z^{\prime}$ boson which contribute to FCNC studied in this work,

\begin{equation}
\mathcal{L}^{K^{0}-\bar{K^{0}}}_{Z'}= \left(  \frac{g}{2C_W}\frac{4S^ {2}_W }{\sqrt{12h_W}} \right)\{ (V^d_L)^{\ast}_{13}(V^d_L)_{23}\}[\bar{d_{2L} }\gamma^\mu d_{1L}]Z^{\prime}_\mu
\label{kk}
\end{equation}

\begin{equation}
\mathcal{L}^{D^{0}-\bar{D^{0}}}_{Z'}= \left(  \frac{g}{2C_W}\frac{4S^ {2}_W }{\sqrt{12h_W}} \right)\{(V^u_L)^{\ast}_{13}(V^u_L)_{23}\}[\bar{u_{2L} }\gamma^\mu u_{1L}]Z^{\prime}_\mu
\label{DD}
\end{equation}

\begin{equation}
\mathcal{L}^{B^{0}_d-\bar{B^{0}_d}}_{Z'}= \left(  \frac{g}{2C_W}\frac{4S^ {2}_W }{\sqrt{12h_W}} \right)\{(V^d_L)^{\ast}_{13}(V^d_L)_{33}\}[\bar{d_{3L} }\gamma^\mu d_{1L}]Z^{\prime}_\mu.
\label{BB}
\end{equation}

Moreover, the interactions among physical standard quarks and $h_1^0$ and $h_2^0$ scalars that contributes to the processes that we study are:

\begin{eqnarray}
\mathcal{L}^{K^{0}-\bar{K^{0}}}_{h_1^0,h_2^0} &=& -\sin_{\beta}\left[\left[K^{D}_{2,1}+(K^{D}_{1,2})^{\ast}\right]\bar{d_2}d_1+\left[K^{D}_{2,1}-(K^{D}_{1,2})^{\ast}\right]\bar{d_2}\gamma_{5}d_1\right]h_1^0\nonumber\\
& &  +\cos_{\beta}\left[\left[K^{D}_{2,1}+(K^{D}_{1,2})^{\ast}\right]\bar{d_2}d_1+\left[K^{D}_{2,1}-(K^{D}_{1,2})^{\ast}\right]\bar{d_2}\gamma_{5}d_1\right]h_2^0
+H.C
\label{kkscalar}
\end{eqnarray}

\begin{eqnarray}
\mathcal{L}^{D^{0}-\bar{D^{0}}}_{h_1^0,h_2^0} &=& -\sin_{\beta}\left[\left[K^{U}_{2,1}+(K^{U}_{1,2})^{\ast}\right]\bar{u_2}u_1+\left[K^{U}_{2,1}-(K^{U}_{1,2})^{\ast}\right]\bar{u_2}\gamma_{5}u_1\right]h_1^0\nonumber\\
& &  +\cos_{\beta}\left[\left[K^{U}_{2,1}+(K^{U}_{1,2})^{\ast}\right]\bar{u_2}u_1+\left[K^{U}_{2,1}-(K^{U}_{1,2})^{\ast}\right]\bar{u_2}\gamma_{5}u_1\right]h_2^0
+H.C
\label{DDscalar}
\end{eqnarray}

\begin{eqnarray}
\mathcal{L}^{B^{0}-\bar{B^{0}}}_{h_1^0,h_2^0} &=& -\sin_{\beta}\left[\left[K^{D}_{3,1}+(K^{D}_{1,3})^{\ast}\right]\bar{d_3}d_1+\left[K^{D}_{3,1}-(K^{D}_{1,3})^{\ast}\right]\bar{d_3}\gamma_{5}d_1\right]h_1^0\nonumber\\
& &  +\cos_{\beta}\left[\left[K^{D}_{3,1}+(K^{D}_{1,3})^{\ast}\right]\bar{d_3}d_1+\left[K^{D}_{3,1}-(K^{D}_{1,3})^{\ast}\right]\bar{d_3}\gamma_{5}d_1\right]h_2^0
+H.C
\label{BBscalar}
\end{eqnarray}

\section{}
\label{appendix2}

Unitary matrices $V^{u,d}_{L,R}$ from \cite{PRDColombia},

\begin{equation}
V^{u}_{L}=V^{u}_{R}=
\begin{pmatrix}
0.89397 & -0.44813 & 0.00046\\
-0.44735 & -0.89233 & 0.06019\\
0.02656 & 0.05401 & 0.99819
\end{pmatrix},
\end{equation}

\begin{equation}
V^{d}_{L}=V^{d}_{R}=
\begin{pmatrix}
0.97361 & -0.22669 & -0.0169663\\
-0.223738 & 0.96825 & 0.0583041\\
-0.0341856 & 0.0536757 & 0.99512
\end{pmatrix}.
\end{equation}

Comparing the mass matrices from Ref.\cite{PRDColombia} with the mass matrices in Eq.\eqref{massmatrixupquarks} we obtain the matrix elements (in Gevs) that enters in the expresions of Eq.\eqref{gu7} and Eq.\eqref{gu8} as follows: $m^{d}_{11} = m^{d}_{12} = m^{d}_{21}=0$, $m^{d}_{13} = 0.127037$, $m^{d}_{22} = -0.0269844$, $m^{d}_{23} = 0.262835$; and of Eq.\eqref{gu7}: $m^{u}_{31} = 4.5398$, $m^{u}_{32} = 9.2318$ $m^{u}_{33} = 170$.

\section{}
\label{appendix3}

Unitary matrices $V^{u,d}_{L,R}$ from Ref.\cite{polones},

\begin{equation}
V^{u}_{L}=
\begin{pmatrix}
0.99987 & 0.0163 & 0.00062\\
-0.0163 & 0.99987 & 0.00064\\
-0.00061 & -0.00064 & 1
\end{pmatrix},
\end{equation}

\begin{equation}
V^{u}_{R}=
\begin{pmatrix}
1 & 0.00003 & 8.15\times10^{-9}\\
-0.00003 & 1 & 4.62\times10^{-6}\\
-8.01\times10^{-9} & -4.62\times10^-{6} & 1
\end{pmatrix}.
\end{equation}

\begin{equation}
V^{d}_{L}=
\begin{pmatrix}
0.97741 & -0.21126 & 0.00624\\
0.21134 & 0.97656 & -0.04079\\
0.00252 & 0.04119 & 0.99915
\end{pmatrix},
\end{equation}

\begin{equation}
V^{d}_{R}=
\begin{pmatrix}
0.99993 & -0.01138 & 7.43\times10^-{6}\\
0.01138 & 0.99993 & -0.00092\\
3.08\times10^{-6} & 0.00924 & 1
\end{pmatrix},
\end{equation}

Comparing the mass matrices from Ref.\cite{polones} with the mass matrices Eq.\eqref{massmatrixupquarks} we obtain the matrix elements (in Mevs) that enters in the expresions of Eq.\eqref{gu7} and Eq.\eqref{gu8} as follows: $m^{d}_{11} = 5.11523$, $m^{d}_{12} = 20.03$, $m^{d}_{13} =  10.5861$, $m^{d}_{21} = 0$, $m^{d}_{22} = 92.9391$, $m^{d}_{23} =172.911$; and in the Eq.\eqref{gu7} $m^{u}_{31} = 0$, $m^{u}_{32} = 0$, $m^{u}_{33} = 172500$.

\section{}
\label{appendix4}

\subsubsection{$K^{0}-\bar{K^{0}}$ System}

From Eq.\eqref{kk}:
\begin{eqnarray}
\ensuremath{\mathcal{L}}^{K_{0}-\bar{K_{0}}}_{Z'\ eff} & = & \frac{4 \sqrt{2} G_F S^4_W}{(3h_W)}\frac{M_{Z}^{2}}{M_{Z^{\prime}}^{2}}[(V_{L}^{d})_{13}^{\ast}(V_{L}^{d})_{23}]^{2}[\bar{d}_{2L}\gamma_{\mu}d_{1L}]^{2},\nonumber \\
\label{effectiveLK}
\end{eqnarray}
then
\begin{eqnarray}
(\Delta m_{K})_{Z^{\prime}} & = & <\bar{K^ {0}}|\ensuremath{\mathcal{L}}^{K_{0}-\bar{K_{0}}}_{Z'\ eff}|K^ {0}>\nonumber \\
&=&\frac{4 \sqrt{2} G_F S^4_W}{(3h_W)}\frac{M_{Z}^{2}}{M_{Z^{\prime}}^{2}}[(V_{L}^{d})_{13}^{\ast}(V_{L}^{d})_{23}]^{2}<\bar{K^ {0}}|(\bar{d_2}d_1)_{V-A}^2|K^ {0}>.
\label{massdifKZlinha}
\end{eqnarray}
From Eq.\eqref{kkscalar}:
\begin{eqnarray}
\ensuremath{\mathcal{L}}^{K_{0}-\bar{K_{0}}}_{h_1^0,h_2^0\ eff} & = & \dfrac{\sin_{\beta}^{2}}{m^{2}_{h_1^0}}\left[\left[K^{D}_{2,1}+(K^{D}_{1,2})^{\ast}\right]^{2}(\bar{d_2}d_1)^{2}+\left[K^{D}_{2,1}-(K^{D}_{1,2})^{\ast}\right]^{2}(\bar{d_2}\gamma_5d_{1})^{2}\right]\nonumber \\
& &+\dfrac{\cos_{\beta}^{2}}{m^{2}_{h_2^0}}\left[\left[K^{D}_{2,1}+(K^{D}_{1,2})^{\ast}\right]^{2}(\bar{d_2}d_1)^{2}+\left[K^{D}_{2,1}-(K^{D}_{1,2})^{\ast}\right]^{2}(\bar{d_2}\gamma_5d_{1})^{2}\right],
\label{effectiveSLK}
\end{eqnarray}
then
\begin{eqnarray}
(\Delta m_{K})_{h_1^0,h_2^0} & = & \lbrace\dfrac{\sin_{\beta}^{2}}{4m^{2}_{h_1^0}}\left[-(K^{D+}_{2,1})^{2}(1-\frac{M_{K}^{2}}{(m_{s}+m_{d})^{2}})+(K^{D-}_{2,1})^{2}(1-11\frac{M_{K}^{2}}{(m_{s}+m_{d})^{2}})\right]\nonumber\\
&+& \dfrac{\cos_{\beta}^{2}}{4m^{2}_{h_2^0}}\left[-(K^{D+}_{2,1})^{2}(1-\frac{M_{K}^{2}}{(m_{s}+m_{d})^{2}})+(K^{D-}_{2,1})^{2}(1-11\frac{M_{K}^{2}}{(m_{s}+m_{d})^{2}})\right]\rbrace \nonumber\\
& & \times<\bar{K^ {0}}|(\bar{d_2}d_1)_{V-A}^2|K^ {0}>.
\label{massdifK-1h1h2}
\end{eqnarray}
The SM contribution is given by
\begin{eqnarray}
(\Delta m_{K})_{SM} & = & \dfrac{G_{f}^{2}m_{c}^{2}}{16\pi^2}[(V_{CKM})_{cd}^{\ast}(V_{CKM})_{cs}]^2<\bar{B^ {0}}|(\bar{d_2}d_1)_{V-A}^2|B^ {0}>.
\end{eqnarray}
Finally
\begin{eqnarray}
(\Delta m_{K})_{RM331} & = & (\Delta m_{K})_{Z^{\prime}}+(\Delta m_{K})_{h_1^0}+(\Delta m_{K})_{h_2^0}+(\Delta m_{K})_{SM} \nonumber\\
 & & \leqslant 3.483 \times 10^{-12}(MeV)
\end{eqnarray}

\subsection{$D^{0}-\bar{D^{0}}$ System}

From Eq.(\ref{DD}) we get:
\begin{eqnarray}
\ensuremath{\mathcal{L}}^{D_{0}-\bar{D_{0}}}_{Z'\ eff} & = & \frac{4 \sqrt{2} G_F S^4_W}{(3h_W)}\frac{M_{Z}^{2}}{M_{Z^{\prime}}^{2}}[(V_{L}^{u})_{13}^{\ast}(V_{L}^{u})_{23}]^{2}[\bar{u}_{2L}\gamma_{\mu}u_{1L}]^{2},\nonumber \\
\label{effectiveLD}
\end{eqnarray}and consequently
\begin{eqnarray}
(\Delta m_{D})_{Z^{\prime}} & = & <\bar{D^ {0}}|\ensuremath{\mathcal{L}}^{D_{0}-\bar{D_{0}}}_{Z'\ eff}|D^ {0}>\nonumber \\
&=&\frac{4 \sqrt{2} G_F S^4_W}{(3h_W)}\frac{M_{Z}^{2}}{M_{Z^{\prime}}^{2}}[(V_{L}^{u})_{13}^{\ast}(V_{L}^{u})_{23}]^{2}<\bar{D^ {0}}|(\bar{u_2}u_1)_{V-A}^2|D^ {0}>.
\label{massdifDZlinha}
\end{eqnarray}

From Eq.(\ref{DDscalar}):

\begin{eqnarray}
\ensuremath{\mathcal{L}}^{D_{0}-\bar{D_{0}}}_{h_1^0,h_2^0\ eff} & = & \dfrac{\sin_{\beta}^{2}}{m^{2}_{h_1^0}}\left[\left[K^{U}_{2,1}+(K^{U}_{1,2})^{\ast}\right]^{2}(\bar{u_2}u_1)^{2}+\left[K^{U}_{2,1}-(K^{U}_{1,2})^{\ast}\right]^{2}(\bar{u_2}\gamma_5u_{1})^{2}\right]\nonumber \\
& &+\dfrac{\cos_{\beta}^{2}}{m^{2}_{h_2^0}}\left[\left[K^{U}_{2,1}+(K^{U}_{1,2})^{\ast}\right]^{2}(\bar{u_2}u_1)^{2}+\left[K^{U}_{2,1}-(K^{U}_{1,2})^{\ast}\right]^{2}(\bar{u_2}\gamma_5u_{1})^{2}\right],
\label{effectiveSLD}
\end{eqnarray}
then
\begin{eqnarray}
(\Delta m_{D})_{h_1^0,h_2^0} & = & <\bar{D^ {0}}|\ensuremath{\mathcal{L}}^{D_{0}-\bar{D_{0}}}_{h_1^0,h_2^0\ eff}|D^ {0}>\nonumber\\
& = & \lbrace\dfrac{\sin_{\beta}^{2}}{4m^{2}_{h_1^0}}\left[-(K^{U+}_{2,1})^{2}(1-\frac{M_{D}^{2}}{(m_{u}+m_{c})^{2}})+(K^{U-}_{2,1})^{2}(1-11\frac{M_{D}^{2}}{(m_{u}+m_{c})^{2}})\right]\nonumber\\
&+& \dfrac{\cos_{\beta}^{2}}{4m^{2}_{h_2^0}}\left[-(K^{U+}_{2,1})^{2}(1-\frac{M_{D}^{2}}{(m_{u}+m_{c})^{2}})+(K^{U-}_{2,1})^{2}(1-11\frac{M_{D}^{2}}{(m_{u}+m_{c})^{2}})\right]\rbrace\nonumber\\
& & \times<\bar{D^ {0}}|(\bar{u_2}u_1)_{V-A}^2|D^ {0}>.
\label{massdifDh-1h1h2}
\end{eqnarray}

Finally
\begin{eqnarray}
(\Delta m_{D})_{RM331} & = & (\Delta m_{D})_{Z^{\prime}}+(\Delta m_{D})_{h_1^0}+(\Delta m_{D})_{h_2^0}\nonumber\\
 & & \leqslant 9.478 \times 10^{-15}(GeV)
\label{uuu} 
\end{eqnarray}

The approach \eqref{uuu} is justified given the very poor knowledge of the SM contributions  to $(\Delta m_{D})$.




\end{document}